# Measuring the Globalization of Knowledge Networks

Caroline S. Wagner[1] and Loet Leydesdorff[2]

[1] SRI International, Arlington, Virginia and the Center for International Science and Technology Policy, George Washington University 1957 E Street NW Washington DC 20052 USA. cswagner@gwu.edu (corresponding author)
[2] Amsterdam School of Communications Studies (ASCoR), University of Amsterdam, Kloveniersburgwal 48, 1018 CX Amsterdam NL, loet@leydesdorff.net; http://www.leydesdorff.net

1. Introduction

This paper presents a method of understanding the growth of global science as resulting from a mechanism of preferential attachment within networks. The paper seeks to contribute to the development of indicators of knowledge creation and transfer by presenting a theory and case study of network structures in science. It is our view that indicator development has suffered from a lack of attention to the theoretical basis for understanding the dynamics of knowledge creation. This lack has been due, in part, to the difficulty of measuring dynamic systems within social organizations. This paper attempts to fill this gap by proposing theory-based indicators of knowledge creation using network theory and analysis. The paper presents a hypothesis about the knowledge system that is explored by analyzing the growth of international collaboration in science.

Scientific research creates knowledge through a process of hypothesis, experimentation, testing, and verification. Funds and infrastructure are committed to experimental research. The processes within the knowledge-creation process produce outputs such as publications and patents. These outputs can be tracked through analysis of the authorship, citations and references within the published materials. A factor limiting analysis has been the lack of direct measures of the process. S&T indicators have tended to rely on inputs to the process (such as funding and staffing) and outcomes of the process (such as publications and patents). The Holy Grail of research assessment, however, has been to look "Inside the Black Box" as Nathan Rosenberg (1974) put it in the title of his book, to view more closely and understand more fully the knowledge creation process as it happens, as well as to understand the dynamics of the system that nurtures the process.

2. Theoretical Basis

Drawing from a variety of intellectual strands coming from organizational studies (Aldrich, 1999; Astley, 1985; Monge & Contractor, 2003; Mohrman, Galbraith & Monge, forthcoming), complexity theory and the study of self-organization (Kauffman, 1993; Holland 1995; Tuomi, 2002), network theory (Barabasi & Albert 1999; Newman 2001) and communications theory (Luhmann 1988; Maturana & Varela 1987, Leydesdorff 2000) we have developed the hypothesis that scientific knowledge emerges along the lines of a emergent, self-organizing system similar to a biological eco-system, but with a different dynamics (also, Wagner & Mohrman, forthcoming). The individual agents in the landscape (in this case, individual scientific researchers and engineers) seek resources, recognition, and reward (Whitely 1984). As they examine the landscape for the most efficient way to obtain these benefits, agents engage in both cooperation and competition in a field of finite resources (Axelrod & Epstein 1994). The agents connect through a series of local searches and connections, as well as through weak links to connect to formerly unknown people or resources. Through their non-linear and complex interactions, the agents create a system that takes on emergent properties of its own. The emergent system then becomes a constraint and a reference point for researchers (Leydesdorff 2001). The combination and recombination of know-how and ideas leads to the creation of knowledge (Weitzman 1993). New knowledge then provides opportunities of preferential attachment in a network.

The dynamics of the knowledge-creating system can be considered and studied as a network. The flow of value from scientific research is best understood through a network lens. This view stems from an understanding of the work of science as fundamentally relational, of advances in knowledge as stemming from knowledge sharing and knowledge combining activities that do not know organizational or, increasingly, disciplinary boundaries, and of value as accruing through use, or application. All this happens in a way that is not anticipated or planned ahead of time—the discoveries emerge from the combination of elements across the landscape as they fit the needs of agents and use available resources.

Science can be considered as a communications system that self-organizes into a complex adaptive system. Complex adaptive systems have been found to share common features (Waldrop 1999). According to Amaral (2004) and others, complex systems have the following features:

- a dynamic internal structure that evolves and interacts in a complex manner;
- emergent behaviors and patterns that are not caused by a single entity in the system but may arise from

simple rules;

- can be described as "open" in the sense that information flows across its boundaries, which in turn are difficult to clearly identify; and
- are composed of complex subsystems;
- are sensitive to initial conditions;
- include both cooperative and competitive behaviors or characteristics.

These characteristics are common among those emergent systems made up of large numbers of active elements that are diverse in both form and capacity (Holland 1995).

The principle of self-organization is used in a number of fields to explain the emergence of complexity in cases where external structures and plans cannot.[3] Emergent self-organization is helpful to us in understanding knowledge systems because order (such as the grouping of researchers into disciplines like physics or biology) cannot be attributed solely to top down or bottom up dynamics.[4] Order appears to arise spontaneously from local interactions of actors even if those actors are not aware of how their actions contribute to a larger order.[5]

Looking at a system in the aggregate (or at some ordered level) reveals striking insights into the dynamics of emergence. At the aggregate, the system shows a diversity of structures that are not random and that could not be anticipated by studying individual actors. The properties of the system at the aggregate level are called "emergent properties" because no single actor in the system is responsible for the properties of the system; neither are the properties constructed according to a plan. They emerged from the collective decisions of hundreds of agents.

Complex adaptive systems can organize into a scale-free state (Barabási and Albert 1999; Katz 2006). In other words, self-organizing systems show an uneven distribution. This means that within a system, some factors will show a high frequency, a large size or a large number of connections and some will show a low frequency, a small size or a few connections. The distribution of features in a system will not be uniform; there is no "average" frequency or level of connection.

The distribution within a complex system can be expressed in graphical terms that show how many times a given size or feature appears in the system. (This can be modeled using the Barabási -Albert model producing power-law or 'scale-free' behavior from simple rules.[6]) In complex systems, the distributions describing the patterns have been shown to follow a particular form called a power law. Power laws have two features that make them different from normal distributions (shown in Figure 1). These differences are that a power law does not peak at its average value like that shown in the figure. Instead, a power law graph starts at a maximum value and then decreases rapidly. Secondly, the rate at which the power law decays is much slower than the decay rate for a normal distribution.[7]

Science is a social network. Social networks can be studied in cases where it is possible to identify a relationship based upon some event or affiliation. Relationships—which can be defined in many different ways, such as friendship, co-employment, or co-authorship—must leave some traceable indicator, or else be self-reported or observed through a survey in order to draw a network. As we noted above, the relational nature of science and technology makes these fields very well suited to network analysis; scientists stake claims to intellectual territory by publishing the results of their work, leaving traces of their communications. Further, intellectual links and collaborative relationships are likely to be acknowledged through citations, co-authorships, or acknowledgements in the text (Melin, 2000; Giles & Council, 2004).

---

[3] Krohn *et al.* 1990; Epstein & Axtell 1996
[4] Kauffman 1995; Cilliers 1998
[5] Holland 1995, 1998
[6] The Barabási -Albert model was introduced in "Emergence of scaling in random networks," *Science* 286 (1999) 509-512.
[7] The emergence of the Internet provided a large network of computerized data that allowed mathematicians to test how well the data fit random graph theory. Faloutsos, Falloutsos, and Faloutsos (1999) suggested that certain 'scale free' power laws for the graph of the Internet fit the data very well. Power-law distributions had been observed considerably earlier: in particular, in 1926 Lotka showed that citations in academic literature follow a power law. Similarly, de Solla Price (1963) showed that the number of scientific papers published by authors followed Lotka's law. Other early investigations into power-law distributions were done by Simon (1955) using Zipf's law (1949), as reported in Bornholdt and Schuster 2003.

The shape of the network—called the typology in network studies—is an emergent property of the organization created by network participants. Typology can determine the usefulness of the network to participants. Smaller networks with high redundancy and clustering can be highly useful if the goal is to develop a common pool of knowledge and a highly specialized language. If the goal of the network is to gain access to new ideas and complementary capabilities, then a network with many loose links to second and third level actors (weak links) can provide more value. The tightly clustered network is more stable and the looser network introduces more variety. Many networks appear to balance stability and variety in different measures.

Within networks, the relative positions of nodes (actors) can be mapped and analyzed. Indeed, network analysts are finding that all kinds of networks (social as well as natural) share common statistical properties of scaling, density, and weak links. The analysis gives insights into the dynamics of the relationships in a network and the ability to offer probabilities about its evolution. For example, one of these analytic constructs is "centrality" which is an assessment of the links that nodes have to other nodes where one node is between many other nodes in the network. These central nodes play an important role in holding the network together, in passing information from node to node (often in a power relationship), and in influencing the nature of interconnection.

The stronger the ties and the denser the set of relationships the more tightly knit is the network and the more it resembles an organization. The denser a network becomes, the more it tends to create its own culture that both enables and constrains interpretations and actions. This includes relations of power and processes of exclusion. Dense networks and strong ties decrease the cost of network operation; they also mean that relationships can become locked-in and perhaps too stable when it comes to allowing new people or ideas to enter the network.

3. International Collaboration as a Network

In another paper (Wagner & Leydesdorff 2005b), we suggested that international collaboration in science is a self-organizing network. The network organizes through a process of preferential attachment. The preferential attachment mechanism is one in which scientists compete and cooperate in order to gain recognition. Part of this involves seeking to work with other well-known colleagues who are desirable as collaborators because they increase the chance that research results will gain attention. The function of preferential attachment operates more freely at the international level because connections are less encumbered by political requirements and social obligations.

The fact that collaborators are willing to engage in a joint activity that takes place across geographic distance is an indicator of the usefulness of the collaboration to those involved; the extra "costs" involved in collaborating at a distance (e.g., the lag in communications time, the cultural and language differences) could otherwise be considered a disincentive to joint activities. Ties made at a distance can be more easily broken because they are freer of social obligations than ties made with physically close colleagues. With physically proximate colleagues, researchers develop a common outlook that results in stability. With colleagues from outside one's regular circle, new ideas are more likely to flow, creating variety. Given that scientific referees place a great deal of value on creativity and innovation, we suggest that international collaboration in science is growing precisely because the weak links create the conditions for variety.

The indicator to support this hypothesis can be found in the growth of international collaborative networks. In an earlier paper (Wagner & Leydesdorff 2005a) we demonstrated that international collaboration grew at a spectacular rate between 1990 and 2000. We also showed that co-authorships at the international level can be shown to have scale free properties, suggesting that the network is a complex self-organizing system. To test this hypothesis further, this paper adds data from 2005 to the earlier study. We expect to find that international collaboration continues to grow quickly, that the network has become denser, and that more countries can be counted as part of the core component of the network.

Data was taken from the Science Citation Index for all articles, reviews and letters in SCI for 2005[8]. In his study entitled Evaluative Bibliometrics, Narin (1976) proposed to count only articles, reviews, and notes as indicators of scientific performance. Braun et al. (1987) vigorously argued in favor of including letters as scientific output.  However, the ISI no longer accounted for the category "notes" after 1995. In the scientific community, therefore, consensus has grown that the sum of the numbers of articles, reviews, and letters can be considered as the indicator of scientific performance of a unit.  These are the items we included in the counts.

---

[8] CD-Rom version.

Data is drawn from all papers appearing in the database for 2005.  Table 1 inventories the dataset for the 2005 data.  Once the data is collected into an occurrence table, it is further analyzed to identify collaborations.  Collaboration is considered to be represented by a co-authorship event.  The country counts were done using integer counting which attributes a count of "1" to each occurrence of co-authorship by an author from one country to another one.  The occurrence of authorship events is then placed into a symmetrical matrix where country names appear on both axes, with the co-occurrence of addresses appearing in the corresponding cell.  With this data, we analyzed the observed as well as a normalized data set.[9]  A co-occurrence table is used to conduct the network analysis using UCInet[10] and Pajek[11] software.

Table 1 Data used for analysis, 2005

| 2005 | Full set | Articles, reviews, letters[12] |
|---|---|---|
| SCI-Expanded | 1,298,566 | 986,831 |
| SCI CD-Rom version | 1,011,366 | 734,750[13] |
| Corporate addresses | 2,048,941 | 1,696,042 |
| Authors | 4,478,809 | 3,301,251 |

The observed network has nodes (in this case, countries) and links (the lines between them representing co-authorship events) that can create the observable and measurable network typology.   Using the full, observed network, Table 2 shows the growth in collaboration in the three years for which data are analyzed.  The number of papers and addresses increased in a step-wise function, while the addresses represented for internationally co-authored articles doubled.  This suggests that many more authors have joined the global system.  The percent of all publications articles that were internationally co-authored also increased from 15.6 percent to 17.4 percent of all publications.

Table 2. Data on international collaboration comparing three years, 1990, 2000, and 2005

| Year | Unique documents in SCI | Addresses in the file | Authors for all records | Internationally co-authored records | Addresses, internationally co-authored records | Percent internationally co-authored documents |
|---|---|---|---|---|---|---|
| 2005 | 986,831 | 1,696,042 | 3,301,251 | 171,402 | 618,928 | 17.4 |
| 2000 | 778,446 | 1,432,401 | 3,060,436 | 121,432 | 398,503 | 15.6 |
| 1990 | 590,841 | 908,783 | 1,866,821 | 51,596 | 147,411 | 8.7 |

Perhaps even more striking than the doubling of addresses at the global level is the change in the network over the period being studied.  Table 3 shows the network analysis for the global network of collaborations for the three years studied.  The number of nodes increased by 20 between 1990 and 2000, but half of this was due to the break-up of the Soviet Union into individual states, most of which began to participate in global science during the decade of the 1990s. The number of countries participating at the global level increased from 192 to 194.  The number of nodes in the network tripled during that same time.  The size of the core component[14] grew from 37 countries to 66 countries in the 15-year period. The network became denser, which means that the authors are increasing the number of collaborations in which they are involved at the global level. The average degree[15] is a measure of the spread of influence across the network: here the increasing measure suggests that influence and power is being spread more and more widely at the global level.

---

[9] The normalization is conducted using the cosine (Hamers et al., 1989; Ahlgren et al, 2003; Klavans & Boyack 2004).
[10] Borgatti, S.P., M.G. Everett, and L.C. Freeman. 1999. *UCINET 6.0 Version 1.00*. Natick: Analytic Technologies.
[11] Batagelj, V, Mrvar, A.. July 2006.  http://vlado.fmf.uni-lj.si/pub/networks/pajek/
[12] The category "notes" was no longer used by the ISI after 1995.
[13] Of these 734,750 records, 722,574 contain one or more corporate addresses.

The average distance across the network is the number of steps it takes to go from any one node in the network to any other node in the network. Here the number in Table 3 shows that the number of steps between nodes is lower than two—a very low number in network terms--and it is dropping. This suggests that the network is becoming more interconnected over time. Any one scientist working at the global level is theoretically within two "handshakes" away from each other.

The diameter is the distance from one side of the network to the other. The diameter of 3 is similar to findings by others of the networks created by scientific collaboration (Newman 2001). A small diameter in a large network is another indication of the connections among the members. It also suggests the possibility of "small worlds" emerging within the network. A small world connection is one made between nodes that are not in the same cluster.

The betweenness measure of the graph as a whole is diminishing over time. This measure shows the distribution of influence across the network. A higher measure suggests concentrations of power and influence, while a diminishing number supports the findings of average degree, that the most influential nodes are becoming less influential as more links and connections are made across the network. This increases the opportunities for new ideas to flow into and through the network from formerly excluded or peripheral members. Both the degree and betweenness measures suggest that the knowledge network is becoming more robust.

The final measure shown in Table 3 is the average clustering coefficient. The data shows a slight rise in the clustering coefficient. Clusters are groups within a network where there are redundant connections. The coefficient measures the likelihood that nodes belong to a cluster. Given the disciplinary nature of much of science, clustering is expected. (In another report, one of us found that collaboration across the European Research Area has a clustering coefficient of higher than 83 percent (Wagner et al., 2005). The slight rise in the clustering coefficient cannot in itself be considered significant, since it is so small, but the fact that the average distance is dropping and the clustering coefficient is rising suggests an even greater level of churn among collaborators than is apparent from looking at the network statistics as a whole.

Table 3. Network analysis of global network of collaborations, 1990, 2000, 2005

| Network index | 1990 network | 2000 network | 2005 network |
| --- | --- | --- | --- |
| Number of nodes | 172 | 192 | 194 |
| Number of links | 1 926 | 3 537 | 9400 |
| Size of core component | 37 | 54 | 66 |
| Network density | 0.1310 | 0.1929 | 0.2511 |
| Average degree | 22.442 | 36.896 | 48.649 |
| Average distance | 1.954 | 1.851 | 1.76 |
| Diameter | 3 | 3 | 3 |
| Graph betweenness | 0.2589 | 0.1617 | 0.144 |
| Average clustering coefficient | 0.784 | 0.787 | 0.789 |

The size of the core component shows a growing number of nations that have the capacity to participate in the global knowledge network. Table 4 (next page) lists the countries appearing in the core component of the global network in 1990, 2000, and 2005. Countries that emerged from within the Soviet Union are highlighted with an asterisk in the column for the year 2000. Countries that are new to the list in 2000 or 2005 appear in bold type.

---

[14] The core component, also called the *k*-core, is the subnetwork within a network where each node has at least *k* neighbors. *K* is determined derived from an analysis based on the size of the entire network.

[15] The Freeman degree centralization measure expresses the degree of inequality or variance in our network as a percentage of that of a perfect star network of the same size. There remains a substantial amount of concentration or centralization in this whole network. That is, the power of individual actors varies rather substantially, and this means that, overall, positional advantages are rather unequally distributed in this network, although this power position is being reduced (Hanneman, R., Riddle, M.. Introduction to Social Network Analysis. http://faculty.ucr.edu/~hanneman/ (9/06).

The number of countries participating at the global level nearly doubles between 1990 and 2005. A surprising feature of this list is the increased participation of African countries in the list by 2005, countries that are generally considered to be scientifically lagging (Wagner et al. 2001). Several Middle Eastern countries also join the list. Two small Eastern European countries fall off the list on 2005 (Latvia, Byelarus) and two small South American countries also fall off the list (Columbia, Uruguay).

4. Conclusion

We expected to find that international collaboration continues to grow quickly, that the network has become denser, and that more countries can be counted as part of the core component of the network. These three expectations about global collaboration in science are satisfied by the 2005 data. This supports our thesis that science at the global level is a complex system operating according to network dynamics. It leads to further suggest that the global system is an emergent pattern that is not caused by a single entity in the system but may arise from simple rules at the level of the researchers themselves. It leads to further expectations about the role of initial conditions in affecting the typology, and in the influence of network dynamics on the evolution of the network over time, that we will explore in future research.

In contrast to the operations of science at the national level--where agencies manage and policy directs investment--no global ministry of science connects people at the international level. The network at the global level self-organizes into a complex knowledge system. This suggests that the spectacular growth in international collaboration may be due more to the dynamics created by the self-interest of individual scientists (agents in network terms) seeking creativity and rewards than it is to any policies made at the national level. The many individual choices of scientists to collaborate may be motivated by the reward structure within science where co-authorships, citations and other forms of professional recognition lead to reputation and access to additional resources.

Given the features that suggest that the global knowledge system is a network further suggests that it should be studied as a network, using network analysis theory and tools. This paper uses traditional bibliometric data but applies new tools drawn from network analysis to understand the growth of the system. It draws from both network theory and network analysis to provide a new way to understand globalization of the knowledge system. As such, it gets closer to understanding the *dynamics* of the system. It further offers new ways of understanding the growth of the knowledge system, the flow of knowledge, and insight into the growth of that system over time. Finally, using network analysis allows new approaches to advising the policy process in encouraging the growth of the knowledge system.

Table 4. Countries in the core component of the global network, 1990, 2000, 2005

| 33 countries form the core set of international co-authorship relations in 1990 | 50 countries form a core set of international co-authorship relations in 2000 | 63 countries form a core set of international co-authorship relations in 2005 |
|---|---|---|
| | **Argentina** | Argentina |
| Australia | Australia | Australia |
| Austria | Austria | Austria |
| Belgium | Belgium | Belgium |
| Brazil | Brazil | Brazil |
| Bulgaria | Bulgaria | Bulgaria |
| | Byelarus* | |
| | | **Burkina-Faso** |
| Canada | Canada | Canada |
| | **Chile** | Chile |
| | **Colombia** | |
| | Croatia* | Croatia |
| Czech Republic | Czech Republic | Czech Republic |
| Denmark | Denmark | Denmark |
| Egypt | Egypt | Egypt |
| | Estonia* | Estonia |
| Finland | Finland | Finland |
| France | France | France |
| German Democratic Republic | | |
| Germany Federal Republic | Germany | Germany |
| Hungary | Hungary | Hungary |
| | | **Iceland** |
| India | India | India |
| | | **Indonesia** |
| | | **Iran** |
| Ireland | Ireland | Ireland |
| Israel | Israel | Israel |
| Italy | Italy | Italy |
| Japan | Japan | Japan |
| | Latvia* | |
| | | **Lebanon** |
| | Lithuania* | Lithuania |
| | | **Malaysia** |
| | | **Malta** |
| | **Mexico** | Mexico |
| | | **Morocco** |
| Netherlands | Netherlands | Netherlands |
| New Zealand | New Zealand | New Zealand |
| | | **Nigeria** |
| Norway | Norway | Norway |
| People's Republic of China | People's Republic of China | People's Republic of China |
| | | **Peru** |

|  |  | **Philippines** |
| --- | --- | --- |
| Poland | Poland | Poland |
| Portugal | Portugal | Portugal |
|  | Romania* | Romania |
| USSR | Russia | Russia |
|  |  | **Serbia-Montenegro** |
|  | **Singapore** | Singapore |
|  | Slovakia* | Slovakia |
|  | Slovenia* | Slovenia |
| South Africa | South Africa | South Africa |
|  | **South Korea** | South Korea |
| Spain | Spain | Spain |
| Sweden | Sweden | Sweden |
| Switzerland | Switzerland | Switzerland |
|  | **Taiwan** | Taiwan |
|  |  | **Thailand** |
|  |  | **Tunisia** |
|  | **Turkey** | Turkey |
|  |  | **Uganda** |
|  | Ukraine* | Ukraine |
| UK | UK | UK |
| USA | USA | USA |
|  | **Uruguay** |  |
|  |  | **Venezuela** |
|  |  | **Viet Nam** |
| Yugoslavia | Yugoslavia | Yugoslavia |